\documentclass[aps,prb,twocolumn,showpacs,amsmath,amssymb]{revtex4}
\usepackage{dcolumn}
\usepackage{graphicx}
\usepackage{epsfig}
\usepackage{bm}
\usepackage{amsmath}
\usepackage{amssymb}

\begin{document}

\title{Graphene-coated holey metal films: tunable molecular sensing\\ by surface
plasmon resonance}

\author{Nicolas Reckinger,$^{1,\dagger}$ Alexandru Vlad,$^{2,\dagger}$ Sorin Melinte,$^{2}$ Jean-Fran\c cois Colomer,$^{1}$ Micha\"{e}l 
Sarrazin$^{1,\dagger}$\footnote{e-mail: michael.sarrazin@unamur.be}}

\affiliation{$^{1}$Research Center in Physics of Matter and Radiation (PMR),
Department of Physics, University of Namur, 61 rue de Bruxelles, B-5000 Namur, Belgium}
\affiliation{$^{2}$Institute of Information, Communication Technologies, Electronics and Applied Mathematics, Electrical Engineering, Universite catholique de 
Louvain, Louvain-la-Neuve, B-1348, Belgium}
\affiliation{$^{\dagger}$These authors contributed equally to this work}

\begin{abstract}
We report on the enhancement of surface plasmon resonances in a holey
bidimensional grating of subwavelength size, drilled in a gold thin film
coated by a graphene sheet. The enhancement originates from the coupling
between charge carriers in graphene and gold surface plasmons. The main
plasmon resonance peak is located around $1.5$ $\mu$m. A lower constraint on the gold-induced doping 
concentration of graphene is specified and the interest of this
architecture for molecular sensing is also highlighted.\\$^\dagger$\textit{%
These authors contributed equally to this work.}
\end{abstract}


\maketitle

Plasmonic devices offer valuable platforms for a wide range of emerging
molecular detection schemes. Among such applications, biosensors are very
promising especially from the point of view of lab-on-a-chip (LOC)
technologies.\cite{LOC} Indeed, plasmon resonances are characterized by both
a strong electric field and a great sensitivity to environmental conditions.
As a consequence, adsorbed species can be detected through the resonance
wavelength shift. In addition, the strong electric field enhancement allows for
surface enhanced Raman spectroscopy, which can be used for single molecule
detection.\cite{sers}

\begin{figure}[t]
\centerline{\ \includegraphics[width=8 cm]{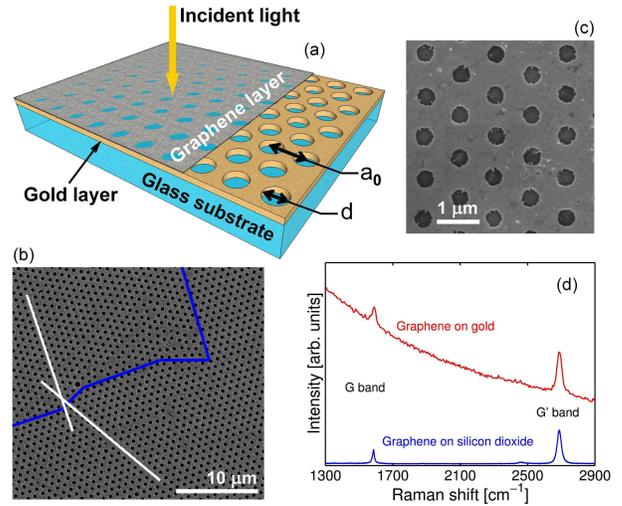}}
\caption{(Color online). (a) Sketch of the device. A microscope glass slide
with a thickness of $1$ mm is coated with a $25$-nm-thick gold film. The
metallic film is perforated by a hexagonal array of holes with a grating
parameter $a_0=980\ \mathrm{nm}$. The hole diameter is $d=405\ \mathrm{nm}$.
The perforated gold film is coated by a graphene sheet. (b) Scanning
electron microscopy (SEM) picture of two domains (delimited by blue lines)
with different orientations (white straight lines) in the holey gold film.
(c) SEM top view of the final device. (d) Raman spectroscopy of the
synthesized graphene layer, transferred on gold and on silicon dioxide.}
\label{fig1}
\end{figure}

Surface plasmons (SPs) require specific conditions to be excited. For
instance, in the Kretschmann configuration, a light beam is totally
internally reflected in a prism on which a metallic film is deposited and
triggers the generation of SPs.\cite{Raether} In a holey metal film, SPs can
be excited by a normal incidence light beam.\cite{Ebbesen} Light is
scattered due to the corrugations and the evanescent diffraction orders can
excite SPs.\cite{Wood} For a metallic layer accommodating an array of holes
with subwavelength size, it is possible to probe SPs by simply measuring the
intensity of the transmitted light. Such a simple configuration is much more
practical in the LOC context and it has been widely studied since the
pioneering work of Ebbesen \textit{et al.} in 1998.\cite{Ebbesen}

Recent theoretical works have shown that doping can induce SP modes in
graphene.\cite{Jablan,Abedinpour,Koppens} Graphene, which appears as a
monoatomic layer made of sp$^2$ carbon atoms in a hexagonal lattice
configuration, presents a plethora of amazing properties.\cite{Geim} In that
context, SPs have been observed for graphene doped by charge transfer from
metal thin films,\cite{Salihoglu,Khomyakov,Giovannetti,Fang} external atoms 
\cite{Shin} or electrostatic gating \cite{Chen,Fei}. It has been also
suggested that SPs could be excited in graphene on a periodically structured
substrate \cite{Zhu,Ferreira} or via regular patterns in graphene.\cite
{Bludov,Nikitin1,Nikitin2,Nikitin3,Popov1,Popov2,Thongrattanasiri} A recent
experimental result showed that graphene SPs can be excited in a Kretschmann
configuration using graphene deposited on a planar gold layer.\cite
{Salihoglu} A different approach \cite{Niu} was used in which SP resonance
tunability was achieved by electromagnetic field coupling between a graphene
sheet and SPs excited in gold nanoparticles.

In the present work, we describe and study an optical device constituted by
a planar hexagonal array of subwavelength-sized holes in a gold thin film
functionalized with a graphene layer. The entire device is built on a glass
substrate. Resonances of the device are measured for various incidence
angles. It is shown that graphene enhances the plasmon resonances and
induces a redshift of the resonance wavelength. This plasmonic device used
as a molecular sensor displays a wavelength shift which is highly sensitive to environmental conditions.

\begin{figure}[t]
\centerline{\ \includegraphics[width=8 cm]{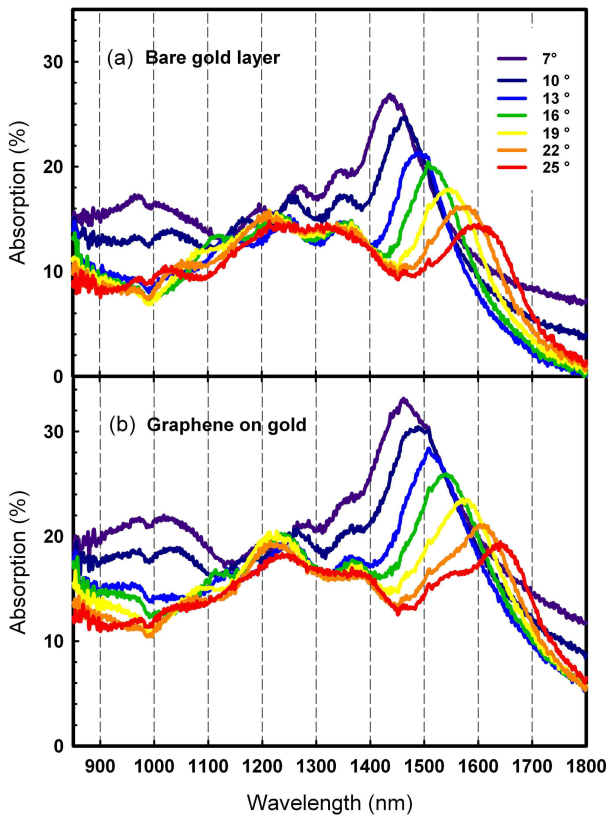}}
\caption{(Color online). Optical characteristics of the device. Absorption
spectra of the uncoated holey gold film (a) and of the graphene-coated
device (b) for various incidence angles.}
\label{fig2}
\end{figure}

Figure 1(a) gives a conceptual illustration of the nanostructured plasmonic
platform. It was fabricated by using colloidal nanosphere lithography.
Briefly, polystyrene spheres ($980$ nm in diameter, Microparticles GmbH)
were deposited on soda-lime glass plates via an interfacial self-assembly
protocol.\cite{vlad1,vlad2} Reactive ion etching using O$_{{2}}$ chemistry
(Oxford Plasmalab, $100$ W RF power, $100$ sccm O$_2$, $25$ mTorr, $12$ min)
was used to reduce the size of the colloids to half the nominal diameter.
Subsequently, $2$ nm of Ti followed by $25$ nm of Au were deposited using
physical vapor deposition. The liftoff was performed using adhesive tape and
ultrasonication in dichloromethane. Figure 1(b) shows a Scanning Electron Microscopy (SEM) top view of the
array made of large periodic crystal-like domains.

Graphene was synthesized by atmospheric pressure chemical vapor deposition
at 1000 $\ensuremath{^\circ}$C on copper foils with methane as carbon source.%
\cite{Li,Wu,Reck} The copper foil was inserted into a quartz reactor inside a
hot-wall furnace. After annealing at 1000 $\ensuremath{^\circ}$C for 30 min
under argon (500 ml/min) and hydrogen (100 ml/min), methane was admitted
(0.5 ml/min) for 15 min to grow graphene. It was next cooled down rapidly
under argon and hydrogen. After spin-coating a polymethyl methacrylate
(PMMA) film over graphene,\cite{dep} the copper foil was etched in aqueous
ammonium persulfate. Next, the PMMA/graphene stack was rinsed in distilled
water and transferred onto the holey gold/glass substrate. The PMMA film was
then removed by soaking into acetone and finally blown dry with nitrogen.
Figure 1(c) shows a SEM top view of the resulting device. In addition, a
reference sample with graphene over planar gold was fabricated. Graphene was
also transferred on silicon dioxide ($300$-nm-thick) for Raman inspection.
The detection of G ($\sim $1590 cm$^{-1}$) and G$^{\prime }$ ($\sim $2690 cm$%
^{-1}$) bands testifies of the presence of graphene (see the Raman spectra
in Fig. 1(d), for graphene over planar gold and silicon dioxide). No
disorder-related D band ($\sim $1350 cm$^{-1}$) is observed, suggestive of
virtually defect-free graphene.

\begin{figure}[t]
\centerline{\ \includegraphics[width=8 cm]{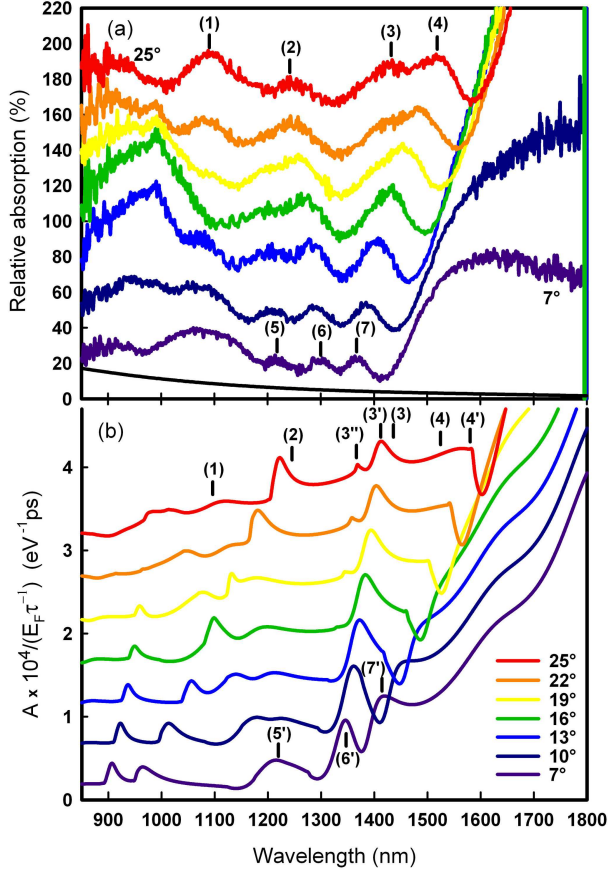}}
\caption{(Color online). (a) Relative absorption of the graphene-coated
versus uncoated gold for various incidence angles. The black curve
represents the relative absorption for a non structured metal film. For the
sake of clarity, each curve is shifted vertically by $25$ units with respect
to the previous one. (b) Numerical simulation of the absorption of the
graphene-coated device for various angles of incidence. For the sake of
clarity, each curve is shifted vertically by $0.5$ units with respect to the
previous one.}
\label{fig3}
\end{figure}

Optical characterization was carried out at various incidence angles by
using an integrating sphere setup mounted on a Perkin-Elmer Lambda 750S
UV/Vis/NIR spectrophotometer. The sample was positioned such that the light
source faces the graphene surface at normal incidence. Figure 2(a) shows the
absorption ($A_\mathrm{B}$) of the bare holey gold film (i.e. without
graphene). The extinction of the graphene-coated holey gold film is shown in
Fig. 2(b) ($A_\mathrm{G} $). The typical absorption peaks denote surface
plasmon polaritons.\cite{Wood} As it can be observed, graphene induces a
redshift of the plasmon resonance peaks and dramatically enhances the
absorption of the device up to $40\%$ (see also Fig. 3(a)). It is noteworthy
that the main plasmon resonance peak is located around $1.5$ $\mu$m, close
to the wavelength range of interest to optoelectronic applications.

The relative absorption $R$, calculated as $R =(A_\mathrm{G}-A_\mathrm{B})/A_%
\mathrm{B}$, is shown in Fig. 3(a). The role of SPs is anticipated from the
relative shift of the peaks in Fig. 3(a) for the considered incidence
angles. In addition, the relative absorption of a graphene layer deposited
on unstructured gold (black curve in Fig. 3(a)) shows a weak enhancement by
contrast with the corrugated gold film. Since corrugated metallic films
allow for SP modes, it is a strong argument for the coupling between SPs on
gold and graphene. To further support this, we have run numerical
simulations based on a homemade code which rests on a rigorous coupled wave
analysis (RCWA) method.\cite{Wood} By considering the frequency $-$ and
position $-$ dependence of the electric field $\mathbf{E}(\omega ,\mathbf{r})
$ inside a given material, the absorbed power is given by:\cite{Jackson}

\begin{equation}
P_{\mathrm{abs}}=-\frac{\varepsilon _0\omega }2\int_V\varepsilon ^{\prime
\prime }(\omega ,\mathbf{r})\left| \mathbf{E}(\omega ,\mathbf{r})\right| ^2%
\mathrm{d}^3r  \label{pow3D}
\end{equation}
where $\varepsilon _0$ is the vacuum permittivity, $\varepsilon ^{\prime
\prime }$ is the imaginary part of the relative permittivity of the medium
(here the graphene layer), and $V$ its volume. This formula can be expressed
according to the real part of the medium conductivity, i.e. $\gamma ^{\prime
}=\varepsilon _0\omega \varepsilon ^{\prime \prime }$. For a
two-dimensional-like medium, such as graphene, we get 
\begin{equation}
P_{\mathrm{abs}}=-\frac 12\int_S\gamma ^{\prime }(\omega ,\mathbf{\rho }%
)\left| \mathbf{E}(\omega ,\mathbf{\rho })\right| ^2\mathrm{d}^2\rho 
\label{pow2D}
\end{equation}
with $S$ the surface, $\rho $ the position vector along the graphene layer,
and where $\gamma ^{\prime }$ is the real part of the graphene conductivity.
\cite{Mermin,Hanson} $\gamma ^{\prime }$ can be restricted to the intraband
term:\cite{Hanson,Hanson2,Jablan}
\begin{equation}
\gamma ^{\prime }=\frac{e^2E_{\mathrm{F}}}{\pi \hbar ^2}\frac{\tau ^{-1}}{%
\left( \omega ^2+\tau ^{-2}\right) }  \label{cond}
\end{equation}
with $\hbar $ the reduced Planck constant, $e$ the electron charge, $E_{%
\mathrm{F}}$ the Fermi level, and $\tau $ the relaxation time. The intraband
approximation is legitimate since no resonance occurs below $900$ nm,
suggesting that $E_{\mathrm{F}}\geq 0.7$ eV,\cite{Bludov,Hanson2} corresponding to
a charge density $n\geq $ 4$\times 10^{13}$ cm$^{-2}$.\cite{Jablan} Albeit large, these
values are in agreement with those achieved in other works.\cite{Khomyakov,Giovannetti,Shin}
\begin{figure}[t]
\centerline{\ \includegraphics[width=8 cm]{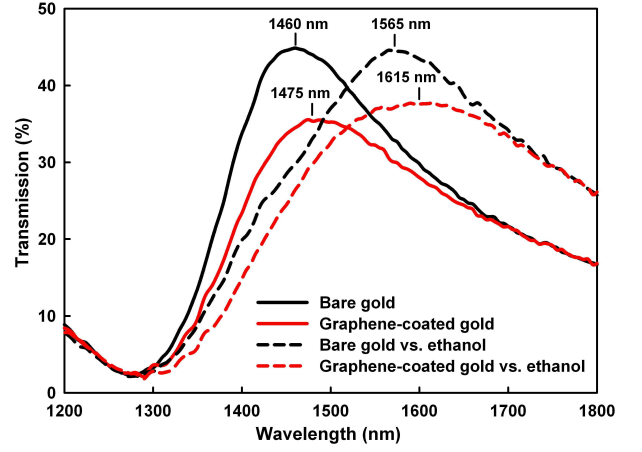}}
\caption{(Color online). Zeroth order transmission through graphene-coated
gold (red curves) and bare gold (black curves) devices. A significant peak
shift occurs when ethanol is deposited on the devices (dashed curves) by
contrast with cases without ethanol (solid curves). Compared to the bare
gold device, for the device coated with graphene, the wavelength-shift is
enhanced by $33\%$. Peaks retrieve their initial positions after few minutes
when ethanol is fully evaporated.}
\label{fig4}
\end{figure}

Let us now consider $\mathbf{E}(\omega ,\mathbf{r})$ as the electric field
of the electromagnetic wave scattered at the air/gold interface without
graphene. We suppose that such an electromagnetic field is not significantly
modified by the graphene layer as a first approximation. In addition, we
suppose that the extrinsic conductivity of graphene due to the gold film can
be considered as uniform, i.e. $\gamma ^{\prime }$ does not depend on $%
\mathbf{\rho }$. The incident power is given by $P_{\mathrm{in}}=(1/2)\sigma
c\varepsilon _0E_{\mathrm{in}}^2$, where $c$ is the speed of light, $\sigma $
is the area of the unit cell of the array, and $E_{\mathrm{in}}$ the electric
field amplitude of the incident wave. Then, the absorption of graphene is
defined as $A=\left| P_{\mathrm{abs}}\right| /P_{\mathrm{in}}$, and assuming
that we work at frequencies $\omega \gg \tau ^{-1}$, we get 
\begin{equation}
A=K_0\lambda ^2\frac 1{E_{\mathrm{in}}^2}\frac 1\sigma \int_S\left| \mathbf{E%
}(\lambda ,\mathbf{\rho })\right| ^2d^2\rho   \label{abs}
\end{equation}
where $\lambda $ is the wavelength ($\lambda =2\pi c/\omega $) and $%
K_0=e^2E_{\mathrm{F}}\tau ^{-1}/(4\pi ^3c^3\varepsilon _0\hbar ^2)$. Except
for $K_0$, which is a mere factor of proportionality, $A$ can be easily
computed numerically using our homemade RCWA code, allowing to propagate the
electromagnetic field on a bare perforated gold layer on glass. The
permittivities of materials are taken from the literature.\cite{Palik} In
Fig. 3(b), we display the calculated quantity $A/E_{\mathrm{F}}\tau ^{-1}$
for different incidence angles. Markers (noted as numerals, from $(1)$ to $%
(7)$, from $(3^{\prime })$ to $(7^{\prime })$, and $(3^{\prime \prime })$)
have been added in Figs. 3(a) and 3(b) for comparison. The global pattern in
Fig. 3(b) matches very well the experimental data shown in Fig. 3(a). This
supports the supposition that the enhancement is due to graphene
conductivity coupled with surface modes, i.e. SPs at the gold/air interface.
We note a few discrepancies, for instance, the peak at wavelength $(3)$
seems to result from two shifted resonances at wavelengths $(3^{\prime })$
and $(3^{\prime \prime }).$ Peaks at wavelengths $(5)$ to $(7)$ also appear 
shifted to wavelengths $(5^{\prime })$ to $(7^{\prime })$. These differences
between experiment and numerical simulations are due to the fact that we
consider a perfectly periodic hole array in our simulation while the real
one is constituted of a set of many crystal-like domains (see the SEM
picture in Fig. 1(b)). In addition, the retroactive role of the graphene
layer on the electric field $\mathbf{E}(\omega ,\mathbf{r})$ is not
considered. The present theoretical results indicate that graphene must be
doped in order to support SPs (here, the doping is extrinsic and provided by
the gold layer). Indeed, if $K_0=0$ (i.e. absence of doping) there is no
absorption. In addition, the intraband approximation is also supported by
the fact that the interband term \cite{Hanson,Hanson2} is unable to
reproduce the experimental patterns shown in Fig. 3(a).

In the following, we show that graphene-coated holey gold films enhance the
wavelength shift against the environment changes proving the versatility of this
configuration for LOC sensing applications. We have analyzed the response of
the devices in the presence of ethanol at the main plasmon resonance around $%
1.5$ $\mu $m. Ethanol is first spread on the graphene side of the device. To
avoid thin film interferences due to a thick ethanol layer, a delay of 30
seconds is set to allow ethanol to partially evaporate. This preserves an
adsorbed layer with a subwavelength thickness. Transmission is then
recorded. Figure 4 displays the response of the devices when in presence of
ethanol. The zeroth order transmissions at normal incidence are plotted
where a typical peak is observed for wavelengths around a SP resonance.\cite
{Wood} The transmission peak for both types of gold samples (graphene-coated
and bare) is redshifted when exposed to ethanol. However, the
graphene-coated sample shows a higher wavelength-shift sensitivity $-$ the
resonance peak shift being $33\%$ greater than for the bare device. After
complete evaporation of ethanol, the peaks retrieve their initial position.
Note that the wavelength shift as well as the transmission decrease (i.e.
the absorption enhancement) are related. Indeed, the surface plasmon wave
vector $k_{\mathrm{sp}}$ must verify $k_{\mathrm{sp}}=\left| \mathbf{k}_{//}+%
\mathbf{g}\right| $, where $\mathbf{k}_{//}$ is the parallel component of
the wave vector of the incident light wave with respect to the interface,
and $\mathbf{g}$ is a vector of the reciprocal lattice. For the smallest non
trivial vector $\mathbf{g}$ and for normal incidence, the wavelength $%
\lambda _{\mathrm{sp}}$ for which a plasmon resonance occurs can be roughly
approximated by:\cite{Jackson,Raether} 
\begin{equation}
\lambda _{\mathrm{sp}}\approx \frac{a_0\sqrt{3}}2Re\left\{ \sqrt{\varepsilon
_d+i\frac{\gamma _d^{\prime }}{\omega \varepsilon _0}}\right\} 
\label{waveres}
\end{equation}
where $\varepsilon _d$ is the permittivity of the dielectric environment which surrounds
the metallic layer and $\gamma _d^{\prime }$ the conductivity of the
dielectric, which depends on the graphene layer conductivity. If $\gamma
^{\prime }$ increases, the optical absorption increases (since $\gamma
^{\prime }=\varepsilon _0\omega \varepsilon ^{\prime \prime }$), but $%
\lambda _{\mathrm{sp}}$ is also shifted. Moreover, the ethanol layer present on the device alters 
the combined permittivity $\varepsilon _d$ and also causes shift and intensity changes.

In summary, nanostructured hybrid graphene-gold architectures enhance the
wavelength shift in plasmonic sensors. The simple graphene-coated metal grating
device allows for a facile optical characterization at normal incidence, without
the need of a complex Kretschmann configuration, greatly improving the
portability of the measurement setup. Moreover, the rich graphene surface
chemistry offers unique functionalization protocols, enlarging further the
sensor's versatility.\cite{Gao}

The authors acknowledge C. N. Santos and B. Hackens for their help with Raman
measurements, as well as F. J. Garcia de Abajo and L. Henrard for useful discussions
and comments. This work was supported by the Belgian Fund for Scientific
Research (F.R.S.-FNRS) under the FRFC contract ''Chemographene'' (convention
No.2.4577.11) and via the FRFC project No.2.4510.11. A.V., S.M. and J.-F.C.
acknowledge F.R.S.-FNRS for financial support. This research used resources
of the Interuniversity Scientific Computing Facility located at the
University of Namur, Belgium, which is supported by the F.R.S.-FNRS under
the convention No.2.4617.07.

\end{document}